\documentclass[twocolumn,aps,prl,superscriptaddress,longbibliography]{revtex4-1}
\usepackage[utf8]{inputenc}
\usepackage{color}
\usepackage{verbatim}
\usepackage{amsmath}
\usepackage{amssymb}
\usepackage{graphicx}
\usepackage[unicode=true,
 bookmarks=false,
 breaklinks=false,pdfborder={0 0 1},colorlinks=true]
 {hyperref}
\hypersetup{
 linkcolor=blue,urlcolor=blue,citecolor=blue}
\usepackage{breakurl}

\makeatletter
\usepackage{mathrsfs}
\usepackage{color}
\usepackage{xcolor}
\usepackage{braket}
\usepackage{tikz}

\usepackage{bm}
\usepackage{bbm}

\makeatother

\begin{document}

\title{Out-of-time-ordered measurements as a probe of quantum dynamics}

\author{Pranjal Bordia}

\affiliation{Fakult\"at f\"ur Physik, Ludwig-Maximillians-Universit\"at M\"unchen, Schellingstr. 4,
80799 Munich, Germany}

\affiliation{Max-Planck-Institut f\"ur Quantenoptik, Hans-Kopfermann-Str. 1, 85748
Garching, Germany}

\author{Fabien Alet}

\affiliation{Laboratoire de Physique Th\'eorique, IRSAMC, Universit\'e de Toulouse,
CNRS, 31062 Toulouse, France}

\author{Pavan Hosur}

\affiliation{Department of Physics, University of Houston, Houston, Texas 77204,
USA}

\affiliation{Texas Center for Superconductivity, Houston, Texas 77204, USA}
\begin{abstract}
Probing the out-of-equilibrium dynamics of quantum matter has gained
renewed interest owing to immense experimental progress in artificial
quantum systems. Dynamical quantum measures such as the growth of
entanglement entropy (EE) and out-of-time ordered correlators (OTOCs)
have been shown, theoretically, to provide great insight by exposing
subtle quantum features invisible to traditional measures such as
mass transport. However, measuring them in experiments requires either
identical copies of the system, an ancilla qubit coupled to the whole
system, or many measurements on a single copy, thereby making scalability
extremely complex and hence, severely limiting their potential. Here,
we introduce an alternate quantity \textendash{} the out-of-time-ordered
measurement (OTOM) \textendash{} which involves measuring a single
observable on a single copy of the system, while retaining the distinctive
features of the OTOCs. We show, theoretically, that OTOMs are closely
related to OTOCs in a doubled system with the same quantum statistical
properties as the original system. Using exact diagonalization, we
numerically simulate classical mass transport, as well as quantum
dynamics through computations of the OTOC, the OTOM, and the EE in
quantum spin chain models in various interesting regimes (including
chaotic and many-body localized systems). Our results demonstrate
that an OTOM can successfully reveal subtle aspects of quantum dynamics
hidden to classical measures, and crucially, provide experimental
access to them. 
\end{abstract}

\maketitle

\section{Introduction}

\label{sec:intro} Quantum phases of matter are generally characterized
by the symmetries broken by their ground or finite temperature states,
the topological properties of the former, or both. These measures
faithfully describe the physics of quantum systems at and near equilibrium.
However, they are not suitable for understanding systems far away
from equilibrium. In this realm, quantum many body systems are better
understood in terms of qualitative features of their dynamics. Classical
measures such as mass transport and quantum measures such as entanglement
entropy (EE) and out-of-time-correlators (OTOCs) \cite{Larkin1969,CalabreseCardy,Maldacena2016}
have been successful in identifying the dynamics in such systems from
a theoretical viewpoint. Seminal examples include the growth of EE after a quench from a high energy state in both chaotic~\cite{KimHuse13}
and non-chaotic systems, particularly many-body localized (MBL) systems~\cite{Znidaric08,Bardarson12}.
The OTOC has also proven very useful for distinguishing between chaotic~\cite{swingle16},
and localized systems~\cite{otoc1,otoc2,otoc3,otoc4}.

Most experiments, however, are designed to measure time-ordered correlation
functions, including classical features such as mass transport. Proposals
for adapting them to measure the aforementioned quantum dynamical
probes exist; however, they are unfeasible even for modest system
sizes, thereby strongly limiting their potential. In particular, current
protocols are rooted in tomography and involve measuring every observable
in the region of interest to reconstruct the density matrix \cite{Tomography16},
or entail measuring the overlap of two wavefunctions by creating two identical copies of the same quantum system \cite{daleyzollareegrowth12,EEMeasurement1,EEMeasurement2,EEMeasurement3,OTOCMeasurement1,OTOCMeasurement2}, interferometrically using a control qubit that couples to the entire system \cite{swingle16} or choosing a special initial state whose corresponding projection operator is a simple observable \cite{swingle16,Meier17}. 
Due to unfavorable scalability, these approaches appear limiting for
a large number of qubits.

In this work, inspired from the OTOCs, we propose an alternative diagnostic,
an out-of-time-ordered measurement (OTOM), to probe and distinguish
three general classes of quantum statistical phases, namely (i) chaotic
(ii) MBL and (iii) Anderson localized (AL).
We also comment on (iv) delocalized integrable systems without disorder
such as freely propagating particles and Bethe integrable systems
(however, we do not study them in detail in this work). All but the
first class are integrable. The second and third classes possess local
conserved quantities that make them integrable, while integrability
in the last class is enforced by non-local conserved quantities.

With regards to experimental accessibility, the OTOM offers several
advantages as compared to OTOCs (since it requires only one copy of
the system) and tomographic approaches (as it involves only one measurement).
Like the OTOCs, it involves reversing the sign of the Hamiltonian
to simulate time reversal \cite{Guanyu2016}, which has been done
in nuclear spin setups \cite{OTOCMeasurement1} and cold atoms experiments
\cite{OTOCMeasurement2} with high precision. Within the scope of
this work, we are not aiming to be able to understand all the detailed
quantum dynamics of the above classes but rather to show that we can
reveal them with the new diagnostic. We comment on the experimental
feasibility of such a protocol towards the end of the work, as well
as on other possible uses of OTOMs.

The plan of the paper is as follows. In Sec.~\ref{sec:def}, we define
the new measure and discuss its relationship with OTOCs. Our analysis
of the OTOM will be performed on spin chain models defined in Sec.~\ref{sec:model},
where details of the numerics are also described. Sec.~\ref{sec:num}
contains the results on the different dynamical behaviors of the OTOM
in the classes of quantum systems investigated. Finally, Sec.~\ref{sec:conc}
offers a discussion of these results and comments on the experimental
protocols to be used to measure the OTOM.

\section{Definitions of dynamical measures}

\label{sec:def}

The OTOM is defined as 
\begin{equation}
\mathcal{M}(t)=\left\langle F^{\dagger}(t)MF(t)\right\rangle 
\end{equation}
where $F(t)=W^{\dagger}(t)V^{\dagger}(0)W(t)V(0)$ (with $W(t)=\exp(iHt)W\exp(-iHt)$)
is an out-of-time-ordered (OTO) operator, $M$ is a simple observable,
and $V$ and $W$ are local \textit{unitary }operators that commute
at $t=0$ \footnote{Unitarity of $W$ and $V$ is not essential; rather, it simplifies
the analysis by allowing us to discard normalization factors associated
with their non-unitarity and is thus, usually assumed.}. Thus, $F(0)=\mathbbm{1}$ and $\mathcal{M}(0)=\left\langle M\right\rangle $.
The expectation value is taken in a suitable state $\rho$ for the
problem of interest, such as a short-range entangled state, an eigenstate
of the Hamiltonian under investigation, or a thermal density matrix.
In this work, we choose $\rho$ to be a short-range entangled pure state, $\rho=\left|\psi\rangle\langle\psi\right|$.

In contrast, the OTOC is defined as: 
\begin{equation}
\mathcal{C}(t)=\left\langle F(t)\right\rangle 
\end{equation}
Both the OTOC and the OTOM involve an OTO operator $F(t)$ and therefore
have many similar properties. For instance, they have many of the
same qualitative features in their time-dependences for
dynamics far from equilibrium in the above phases. However, they have
important differences, that we will exploit in this work to demonstrate
potential advantages of OTOMs in experiments.

Generally, OTOCs can be measured experimentally by applying $F(t)$
on $|\psi\rangle$ to obtain $|\phi(t)\rangle=F(t)|\psi\rangle$.
$F(t)$ contains both forward and backward time-propagation operators,
$e^{\pm iHt}$; in practice, the latter is implemented in experiments
by negating the Hamiltonian. One way to measure a generic OTOC, $\mathcal{C}(t)$,
is by measuring the overlap of $|\phi(t)\rangle$ with a second copy
of $|\psi\rangle$. Creating two copies of the system, however, becomes
prohibitively difficult for large systems. In contrast, OTOM is simply
the expectation value of a local observable $M$ in $|\phi(t)\rangle$,
and does not require any qubit besides the original system. Note that
this procedure is distinct from measuring the dynamical time evolution
of a local observable $M$ after a quench; the latter is known to
not reveal the dynamics captured by EE/OTOCs.

\paragraph{{OTOM vs OTOC:}}

To see how $\mathcal{M}(t)$ is related to $\mathcal{C}(t)$, let
us first consider a trace of the form $I=\mbox{tr}(ABCD)$ for some
operators $A\dots D$. Now, any operator can be thought of as a state
in a doubled Hilbert space: 
\begin{eqnarray}
A & \equiv & \sum_{ij}a_{ij}\left|i\rangle\langle j\right|\to\sum_{ij}a_{ij}|i\rangle|j\rangle\equiv|A\rangle\\
\langle A| & \equiv & \left(\sum_{ij}a_{ij}|i\rangle|j\rangle\right)^{\dagger}=\sum_{ij}a_{ij}^{*}\langle i|\langle j|
\end{eqnarray}
If the Hamiltonian of the original system is $H$, then the doubled system is governed by $\tilde{H}=H\otimes-H$, since kets
in the second copy map to bras in the original system. Then, we can
think of $A$ and $C$ as states on the two copies, and think of $D\otimes\mathbbm{1}$
and $\mathbbm{1}\otimes B$ as operators acting on the first and second
copy, respectively. Note that this doubling is a purely theoretical
construct and is unrelated to the duplication of the experimental
system needed for measuring $\mathcal{C}(t)$. Thus, we have 
\begin{equation}
I=\mbox{tr}(ABCD)=\left\langle A^{*}\left|D^{T}\otimes B\right|C^{T}\right\rangle .
\end{equation}
Choosing $A=\rho^{*}$,
$B=F^{T}(t)$, $C=M^{T}$ and $D=F^{*}(t)$, 
\begin{eqnarray}
\mathcal{M}(t)=\mathcal{M}^{*}(t) & =\mbox{tr}\left[\rho^{*}F^{T}(t)M^{*}F^{*}(t)\right]\\
 & =\left\langle \rho\left|F^{\dagger}(t)\otimes F^{T}(t)\right|M\right\rangle 
\end{eqnarray}
Equivalently, 
\begin{equation}
\mathcal{M}(t)=\left\langle \rho\left|\tilde{V}^{\dagger}(0)\tilde{W}^{\dagger}(t)\tilde{V}(0)\tilde{W}(t)\right|M\right\rangle 
\end{equation}
where $\tilde{V}=V^{T}\otimes V^{*}$, $\tilde{W}=W^{T}\otimes W^{*}$
and time evolution is given by $\tilde{U}(t)=e^{iHt}\otimes e^{-iHt}$,
i.e., $\tilde{W}(t)=\tilde{U}(t)\tilde{W}(0)\tilde{U}^{\dagger}(t)$.
Thus, $\mathcal{M}(t)$ is similar to an OTOC constructed from local
operators $\tilde{V}$ and $\tilde{W}$ and the Hamiltonian $\tilde{H}$,
albeit with two differences: (i) it is defined on two copies of the
same system (it is important to note that the doubled system has the
same localization properties as the single copy), and (ii) the matrix
element of the OTO operator is evaluated between two different states
$|\rho\rangle$ and $|M\rangle$, whereas standard OTOCs are an expectation
value of an OTO operator in a given state.

It is not known generally how off-diagonal matrix elements of OTO
operators behave in various classes of systems. However, if the two
states have a large overlap, we can expect them to qualitatively mimic
traditional OTOCs which are expectation values. Thus, we need: 
\begin{equation}
\left|\left\langle \rho|M\right\rangle \right|\equiv\frac{\left|\text{tr}(\rho M)\right|}{\sqrt{\text{tr}\rho^{2}}\sqrt{\text{tr}M^{2}}}
\end{equation}
In a Hilbert space of dimension $d$, the typical magnitude of the
overlap between two vectors is $1/\sqrt{d}$. In this paper, we work
with a spin-1/2 chain of length $L$ for all our numerical calculations,
and choose $M=\frac{1}{L}\sum_{i}(-1)^{i}\sigma_{i}^{z}$  and $|\psi\rangle=|\uparrow\downarrow\uparrow\downarrow\dots\rangle$. Then $\left\langle \rho|M\right\rangle =1/(\sqrt{L}2^{L/2})\gg1/\sqrt{d}$
where $d=4^{L}$ for the doubled system. Thus, $\left\langle \rho|M\right\rangle $
is ``large'' and $\mathcal{M}(t)$ is expected to behave like a
traditional OTOC in many ways. Indeed, when $M=\rho=|\psi\rangle\langle\psi|$ (so that $\langle\rho |M\rangle=1$), then $\mathcal M(t) = |\mathcal C(t)|^2$, and we recover the protocol for measuring OTOCs proposed theoretically ~\cite{swingle16} and implemented experimentally for a small system ~\cite{Meier17}.

We note that an exception occurs when $M$ is a global conserved quantity.
In the doubled system, this corresponds to $|M\rangle$ being an eigenstate
of $\tilde{H}$. Then, $M$ is a sum of an extensive number of local
terms, and satisfies $[M,H]=0$, while $[M,W]$ and $[M,V]$ contain
$O(1)$ terms each. Thus, $W$ and $V$ can be moved through $M$
in $F(t)$ at the cost of $O(1)$ extra terms, resulting in 
\begin{equation}
\frac{\mathcal{M}(t)}{\mathcal{M}(0)}\to1+O(1/\text{volume})\,\,\,\forall t
\end{equation}
Thus, the time-dependence of $\mathcal{M}(t)$ is suppressed by the
inverse volume of the system when $M$ is a global conserved quantity,
so $\mathcal{M}(t)$ stays pinned at its $t=0$ value for all times
in the thermodynamic limit \footnote{Without loss of generality, we can make $\mathcal{M}(0)\neq0$ by
shifting $M$ by a term proportional to identity, which has the same
scaling with volume as a typical eigenvalue of $M$.}. This behaviour is strikingly different from that of OTOCs in chaotic,
MBL and AL phases, where, as we shall see in the following sections,
the OTOCs decay exponentially, as a power law, and undergo a short-time
evolution before saturating, respectively. However, we will not be
concerned with such special choices of $M$ in this work, since global
conserved quantities can invariably be either factored out trivially
from the analysis, or are too complex to measure experimentally.

Clearly, our specific choices of $M$, $V$, $W$ and $|\psi\rangle$
are not necessary for the above qualitative statements as long as
$V$ and $W$ are local and unitary, $M$ is the sum of an extensive
number of local terms and $\left\langle \psi\left|M\right|\psi\right\rangle $
is not exponentially small itself. Moreover, we expect these results
to generalize to mixed states with short-range entanglement, such
as the infinite temperature state, instead of the pure state $|\psi\rangle$,
and to extend to other choices of operators and states in generic
quantum systems.

\section{Model and methods}

\label{sec:model}

To probe various dynamical behaviors, we consider the spin-1/2 XXZ
chain in a random magnetic field: 
\[
H(\Delta,h)=J\sum_{i=1}^{L}\sigma_{i}^{x}\sigma_{i+1}^{x}+\sigma_{i}^{y}\sigma_{i+1}^{y}+\Delta\sigma_{i}^{z}\sigma_{i+1}^{z}-\sum_{i}h_{i}\sigma_{i}^{z}
\]
We use open boundary conditions and set $J=1$. The random fields
$h_{i}$ are taken from a box distribution $[-h,h]$. This model offers
the advantage of testing different regimes based on the values
of its parameters. The special value $\Delta=0$ corresponds
to a free-fermion model. In presence of disorder, this model is celebrated
to host a MBL phase at large enough disorder $h>h_{c}$ for $\Delta\neq0$,
while the low-disorder phase $h\lesssim h_{c}$ is an ergodic, non-integrable,
system. In the Heisenberg case ($\Delta=1$), the critical disorder
is estimated to be $h_{c}\simeq7.5$ (for eigenstates near the middle of the spectrum)~\cite{Luitz15}. When $\Delta=0$, the model
maps to an Anderson-model which is in a localized phase for
any $h\neq0$.

We choose the OTOM operator $M=\frac{1}{L}\sum_{i}(-1)^{i}\sigma_{i}^{z}$
(staggered magnetization), and the initial state $|\psi\rangle$ as
$|\uparrow\downarrow\uparrow\downarrow\dots\rangle$ for all the numerical
simulations. We use Pauli spin operators for the local unitary operators,
typically $V=\sigma_{i}^{x},W=\sigma_{j}^{x}$ or $V=\sigma_{i}^{z},W=\sigma_{j}^{z}$.

In the numerics, we use full diagonalization in the $\sigma^{z}=\sum_{i}\sigma_{i}^{z}=0$
sector (as $\sigma^{z}$ commutes with $H(\Delta,h)$) on chains of
length up to $L=16$, to probe very long time scales. In some cases,
we also checked our results on larger systems using a Krylov propagation
technique~\cite{Nauts}, albeit on smaller time-scales. We average
quantities over 200 to 1000 disorder realizations for each parameter
set. All times are measured in units of $1/J$, which is set to $1$.

The OTOM $\mathcal{M}(t)$, being an expectation value of an observable,
is real, while the OTOC is in general a complex number. There are
cases however where its imaginary part vanishes (for our choice of
the initial state, this happens for $V=\sigma_{i}^{z},W=\sigma_{j}^{z}$).
When the OTOC is complex, we average its modulus $|\mathcal{C}(t)|$
over disorder, while we average over its real part when it is purely real.

Besides these out-of-time order measurements, to highlight the difference
between quantum and classical measures, we study two properties
of the time-evolved state $|\psi(t)\rangle=\exp(-iHt)|\psi\rangle$: the time-ordered spin imbalance $\mathcal{I}(t)=\langle\psi(t)|M|\psi(t)\rangle$
and the half-system EE, $S(t)=\mbox{tr}(\rho_{A}(t)\ln\rho_{A}(t))$,
where $\rho_{A}(t)=\mbox{tr}_{\bar{A}}|\psi(t)\rangle\langle\psi(t)|$
is the reduced density matrix obtained by tracing out the degrees
of freedom in the other half part $\bar{A}$ of the system.

\section{Numerical Results}

\label{sec:num}

\subsection{Local mass transport (imbalance)}

\begin{figure}
\centering \includegraphics[width=1\linewidth]{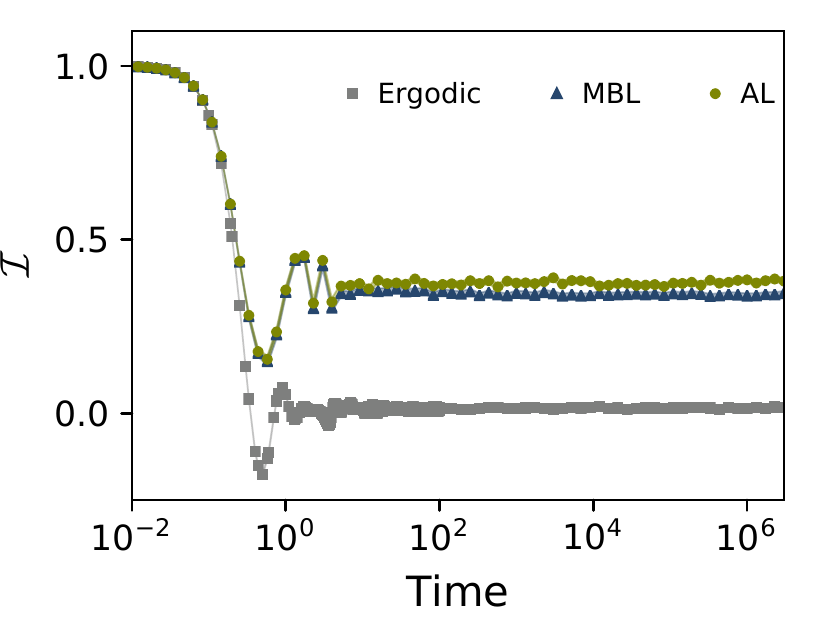}
\caption{\textbf{Local mass transport:} Imbalance $\mathcal{I}$ (with respect
to Néel state) after a quench versus time (log scale), for a $L=16$
spin chain in three different regimes: AL (for $\Delta=0,h=5$),
MBL ($\Delta=0.2,h=5$) and ergodic ($\Delta=1,h=0.5$).}
\label{fig:imbalance} 
\end{figure}

In ergodic systems, we expect local densities to relax exponentially
fast, as a spin density was shown to do when starting from a classical
staggered state~\cite{Barmettler09}. In contrast, this relaxation
is incomplete in localized systems, and the spin density settles to
a non-zero value both in the presence and the absence of interactions.
Fig. \ref{fig:imbalance} shows results for the dynamics
of the imbalance $\mathcal{I}(t)$ in three different regimes of
the XXZ chain. Mass transport is measured experimentally in a number of systems, and is often taken as a probe of local thermalization~\cite{Trotzky12,Pertot14,Schreiber15}.

\subsection{Quantum phase dynamics}

This section contains numerical results for the time evolution
of the above measures for a non-integrable system with chaotic transport
and for MBL/AL systems with absent particle transport. These examples
are chosen to illustrate the important features of both classical
and quantum dynamics. In each case, it appears that the
OTOM can reveal the underlying quantum dynamics.

\subsubsection{Ergodic systems}

\begin{figure}
\centering \includegraphics[width=1\linewidth]{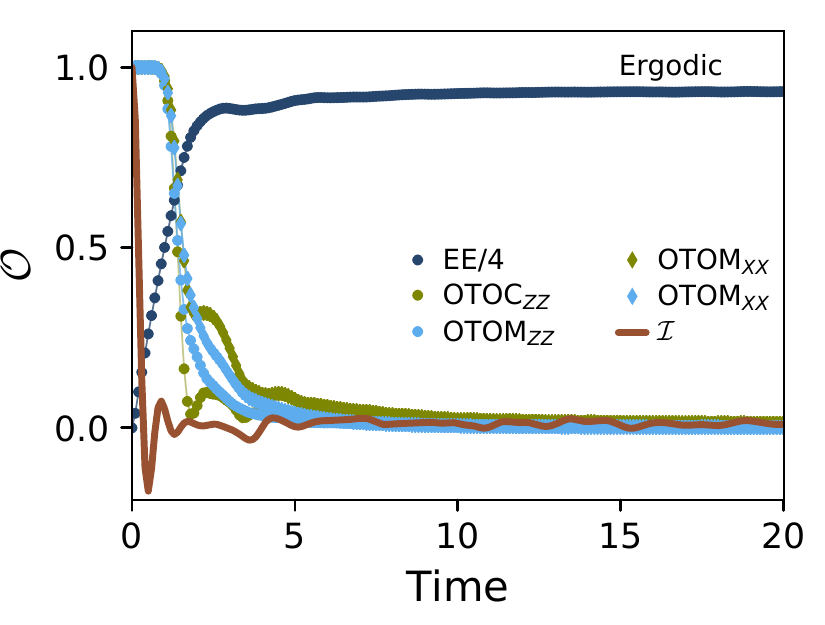} \caption{\textbf{Dynamics in a chaotic system:} The OTOC $|\mathcal{C}(t)|$
as well as the OTOM $|\mathcal{M}(t)|$ relax to zero within a couple
of tunneling times, the time-ordered imbalance $\mathcal{I}(t)$ vanishes
even faster, and the EE $S(t)$ very rapidly saturates
to an extensive value. For the OTOC/OTOMs, two sets of results are
displayed using different local unitary operators: the OTOM/OTOC denoted
XX (ZZ) correspond to $V=\sigma_{5}^{x}$, $W=\sigma_{12}^{x}$ ($V=\sigma_{5}^{z}$,
$W=\sigma_{12}^{z}$). Other parameters are $L=16$, $h=0.5$, $\Delta=1$.
Note the linear time scale, as well as the scaling factor for the
EE.}
\label{fig:chaotic} 
\end{figure}

In a non-integrable system with mass transport (spin transport in
the spin chains considered), we expect time-ordered local observables
to quickly relax to their long-time expectations values, which is
zero in the case of the imbalance $\mathcal{I}$. At the same time,
the OTOCs are also expected to relax exponentially quickly to zero~\cite{Maldacena2016},
and we anticipate the same for the OTOM. This is indeed what we find,
as shown in Fig.~\ref{fig:chaotic}. As shown in the figure, all
operators relax rapidly to zero within a few spin-flip times, and
this holds for different choices of the local operators $V$ and $W$
for the OTOC/OTOM. We note that in such a system, the EE grows very fast (ballistically) and saturates to a volume
law in the subsystem size~\cite{KimHuse13}.

\subsubsection{Localized systems}

Perhaps the clearest distinction of quantum phase dynamics can be
observed in dynamics of localized systems. Localization prohibits
particle/spin transport completely~\cite{Anderson58,Basko06}. Upon
performing a quench of the Hamiltonian, it results in a local memory
of initial density whether or not interactions are present in the
system (see Fig.~\ref{fig:imbalance}). However, non-local quantities
such as EE show distinct behaviors in the presence
or absence of interactions due to the involved decoherence (entanglement)
of the particle phases between nearby sites~\cite{Znidaric08,Bardarson12,Serbyn13}.
OTOCs readily capture this nature of the quantum dynamics as well~\cite{otoc1,otoc2,otoc3,otoc4,otoc5}.

\begin{figure}
\centering \includegraphics[width=1\linewidth]{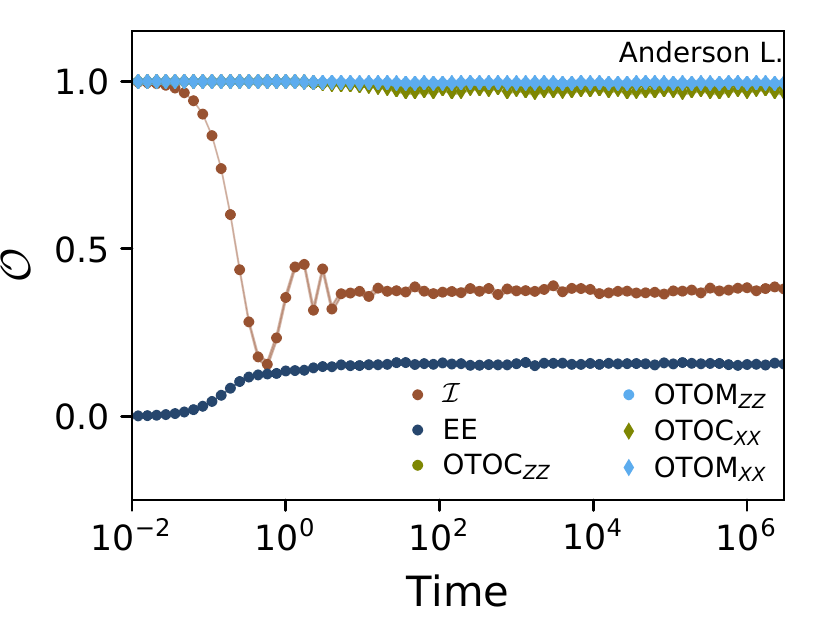} \caption{\textbf{Dynamics in an Anderson localized system:} The time-ordered
imbalance $\mathcal{I}(t)$ relaxes to non-zero value due to absence
of transport (same data as Fig.~\ref{fig:imbalance}) \textit{and}
the OTOC/OTOM are both close to unity and overlapping. At the same
time, the EE $S(t)$ quickly saturates to a finite
small value. In such a well localized system, there is almost no intermediate
dynamics. Parameters for simulations: $L=16$, $\Delta=0$, $h=5$.
For the OTOC/OTOM, we consider operators $V=\sigma_{5}^{x}$, $W=\sigma_{12}^{x}$
(denoted as XX) as well as $V=\sigma_{5}^{z}$, $W=\sigma_{12}^{z}$
(denoted as ZZ) \textendash{} data for these two different choices
are not distinguishable on this scale. }
\label{fig:al} 
\end{figure}

For non-interacting AL systems, particle phases do
not entangle with one another. The time evolution of the various quantities
is shown in Fig.~\ref{fig:al}. The EE shows no
further time dynamics after an initial increase (due to a finite localization
length). Such quickly saturating behavior and absence of dynamics
at longer times can also be captured by OTOCs~\cite{otoc1,otoc2,otoc3,otoc4}.
They equal unity in the absence of any time evolution (by definition),
and deviate only very slightly from this value with time. More precisely,
while their exact time dependence depends on the local spins perturbed
by the operators $V$ and $W$ and the strength of the disorder, importantly,
they show no time dynamics after an initial resettling (of the order
of $~50$ tunneling times for the parameters of Fig.~\ref{fig:al}),
strongly resembling the dynamics of EE. The time
evolution of OTOM for an AL system also displays the
same qualitative features: a tiny reduction from unity and no
marked time dependence, even up to very long times. This, thus, also
provides evidence for the absence of further quantum phase dynamics
in such systems, similar to the analysis of the dynamics of entanglement
(a non-local probe). The absence of decay is a generic feature, as
can be observed for different choices of local operators $V$ and
$W$ in Fig.~\ref{fig:al}. However, one may also argue that the
same kind of information is carried by the (absence of) dynamics in
a time-ordered quantity such as the normal imbalance (also displayed
in Fig.~\ref{fig:al}), a very local `classical' probe. Hence, to clearly show that the two observables capture qualitatively
different information, it is crucial to look at the interacting generalization
of AL systems.

\begin{figure}
\centering \includegraphics[width=1\linewidth]{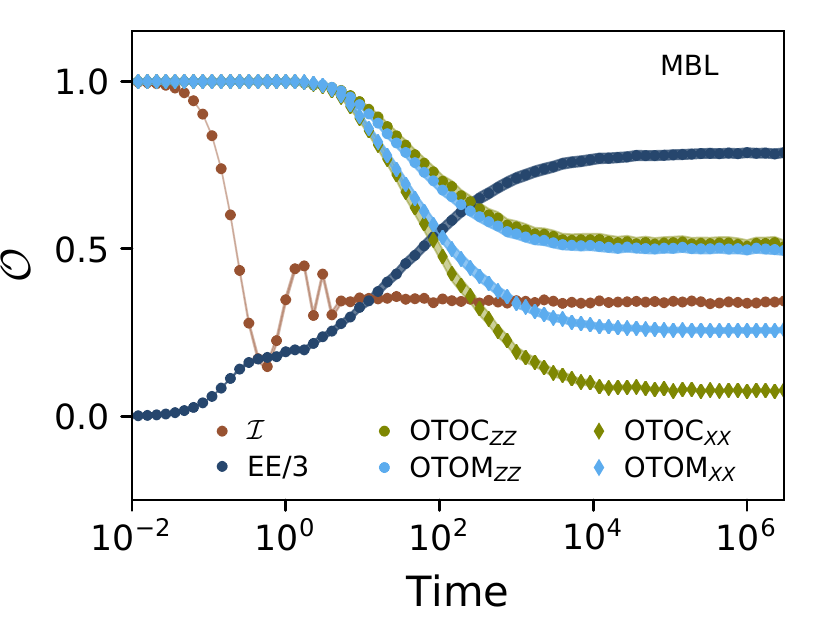} \caption{\textbf{Dynamics in a many-body localized system:} Similar to AL,
the imbalance $\mathcal{I}$ relaxes to a non-zero value on a relatively
short time scale (same data as Fig.~\ref{fig:imbalance}). On the
other hand, the EE $S(t)$ shows a slow logarithmic
growth and OTOCs show an apparent slow power-law decay, revealing the quantum phase dynamics at
time much longer than the density relaxation time. This slow dynamics
is also revealed in OTOMs, which display the same qualitative behavior
as OTOCs. Note the lin-log scale for the main panel, the log-log scale
for the inset, and the scaling factor for the EE.
Parameters for simulations: $L=16$, $\Delta=0.2$, $h=5$. For the
OTOC/OTOM, we consider operators $V=\sigma_{5}^{x}$, $W=\sigma_{12}^{x}$
(denoted as XX) and $V=\sigma_{5}^{z}$, $W=\sigma_{12}^{z}$ (denoted
as ZZ) . The exact long-time limit of OTOC/OTOM depends on the operators
chosen for $V$ and $W$.}
\label{fig:mbl} 
\end{figure}

The interacting case of MBL systems offers quite a different situation,
due to the slow entangling of quantum phases even in the absence of
transport. The numerical calculations for this case are shown in Fig.~\ref{fig:mbl}.
In such systems, the EE indeed shows time dynamics
despite the complete absence of transport, featuring a slow logarithmic
growth~\cite{Znidaric08,Bardarson12,Serbyn13} to reach an extensive
value (which is smaller than the one reached in the ergodic phase~\cite{Serbyn13}).
At the same time, recent results have shown that time dynamics in
MBL systems can also be captured by OTOCs, which have been shown to
exhibit a slow power-law like relaxation~\cite{otoc1,otoc2,otoc3,otoc4}.
We also find that the OTOM is capable of capturing the time dynamics
in the MBL phase: it shows a slow relaxation similar to the OTOC behavior,
albeit it does not reach the same long-time value. This can be understood
as the long-time limit of the OTOC depends on the overlap of $V$
and $W$ with the local integral of motions inherent to the MBL phase~\cite{otoc4}
and we consequently expect that the long-time limit of the OTOM depends
on analogous overlaps between $V$, $W$ and $M$, resulting in different
values.

This observation of slow relaxation at intermediate times is in strong
contrast with the behavior of the time-ordered imbalance, which again
shows no time dynamics at these timescales of slow dephasing. 

\section{{Discussions and conclusions}}

\label{sec:conc}

We have demonstrated that out-of-time ordered measurements OTOMs carry
the same information as their correlator counterpart OTOC for three
different physical situations: ergodic, AL and MBL systems. We propose a concrete measure which we believe
will advance the use of genuinely interesting quantum probes
in experiments. Thanks to its scalability, one main application of
our probe is to allow experimental access to higher dimensional systems
and in particular, to two dimensional systems debated to have an MBL
phase~\cite{2DMBLLocal}. While reversing the sign of the Hamiltonian
is a hard task, each term of the Hamiltonian has
been sign-reversed in several experiments, including hopping (with
a drive), interactions (using Feshbach resonance) and disorder (using
a phase shift of the quasi-periodic term). Additionally, this opens
the door to measuring quantum dynamics in systems where the initial
state cannot be duplicated, such as superconducting qubits, trapped
ions and NV-centers.

On the theoretical front, our work naturally opens up several avenues
for future studies. Firstly, given that the OTOM is an off-diagonal
OTOC in a doubled system, one can ask, ``what features of quantum
dynamics are captured by off-diagonal OTOCs that are missed by the
usual diagonal OTOCs'' and vice versa? Secondly, one can try to exploit
the freedom in choosing the OTOM operator $M$ to study aspects of
quantum dynamics not studied here. One direct application would be
to explore the SYK model \cite{KitaevTalk,Sachdev1993SYK,MaldacenaStanfordSYK,Polchinski2016}
and the debated Griffiths-phase before the ergodic-MBL transition~\cite{agarwal_rare-region_2017,luitz_ergodic_2017,prelovsek_density_2017}.
Another possible continuation is to consider delocalized systems which
are integrable, such as systems of free particles or systems soluble
by the Bethe ansatz, as they necessarily have non-trivial global conserved
quantities (i.e., conserved quantities in addition to trivial ones
such as total charge or spin). We noted earlier that OTOMs behave
differently from OTOCs when $M$ is chosen to be a global conserved
quantity. It would be an interesting extension of our work to compare and contrast
OTOCs and OTOMs in these phases in detail. We expect that OTOCs, as
well as OTOMs derived from non-conserved quantities would exhibit
chaotic (oscillatory) time dependence in Bethe-ansatz soluble (free particle)
systems, while OTOMs based on global conserved quantities would show
a suppression of any time-independence at all with system size. Finally,
unlike ordinary expectation values, OTOMs are able to distinguish
between local and global conserved quantities, since only OTOMs based
on the latter are expected to remain pinned at their $t=0$ values
in the thermodynamic limit. Naively, local conserved quantities are
expected to reduce chaos in a system more than global ones are; OTOMs
might be ideally suited for capturing this difference. We leave these
open questions for future investigations.

\begin{acknowledgments}
This work benefited from the support of the project THERMOLOC ANR-16-CE30-0023-02
of the French National Research Agency (ANR) and by the French Programme
Investissements d'Avenir under the program ANR-11-IDEX-0002-02, reference
ANR-10-LABX-0037-NEXT. FA acknowledges PRACE for awarding access to
HLRS's Hazel Hen computer based in Stuttgart, Germany under grant
number 2016153659, as well as the use of HPC resources from GENCI
(grant x2017050225) and CALMIP (grant 2017-P0677). PH was supported
by the Division of Research, the Department of Physics and the College
of Natural Sciences and Mathematics at the University of Houston. 
\end{acknowledgments}

\bibliography{QuantumDynamic_arxiv_v1}

\end{document}